# Laboratorial radiative shocks with multiple parameters and first quantifying verifications to core-collapse supernovae

Lu Zhang[a], Jianhua Zheng[a], Zhenghua Yang[a], Tianming Song[a], Shuai Zhang[a], Tong Liu[b, *], Yunfeng Wei[b], Longyu Kuang[a, *], Longfei Jing[a], Zhiwei Lin[a], Liling Li[a], Hang Li[a], Jinhua Zheng[a], Pin Yang[a], Yuxue Zhang[a], Zhiyu Zhang[a], Yang Zhao[a], Zhibing He[a], Ping Li[a], Dong Yang[a], Jiamin Yang[a], Zongqing Zhao[a], Yongkun Ding[c, *]

[a] Research Center of Laser Fusion, China Academy of Engineering Physics, Mianyang 621900, China

[b] Department of Astronomy, Xiamen University, Xiamen, Fujian 361005, China

[c] Institute of Applied Physics and Computational Mathematics, Beijing 100088, China

* Corresponding authors.

E-mail addresses: ding-yk@vip.sina.com (Yongkun Ding), tongliu@xmu.edu.cn (Tong Liu) and kuangly0402@sina.com (Longyu Kuang)

A radiative shock (RS) is one in which the density and temperature structures are affected by radiation from the shock-heated matter. RS plays a special role in astrophysics as it nontrivially combines both hydrodynamics and radiation physics. In most astrophysical shocks, the temperature and density conditions lead to strong emission, with radiation thus playing a major role therein. Various RS structures can be implied for numerous astrophysical objects, such as supernova explosions, stellar interiors [1], stellar winds, star formation, black hole accretion disks [2], accreting neutron stars [3], and gamma-ray bursts [4]. In particular, RS exists in the blast waves of core-collapse supernovae (CCSNe), where the radiation pressure in matter is larger than the thermal one. On the basis of multi-messenger supernova observations, their explosion model and particle acceleration mechanisms have been built, and the characteristics of progenitor stars can be further constrained. However, entire supernova explosions are very difficult to model numerically because of the different spatial and temporal length scales involved and the controversial neutrino-driven mechanism [5]. Direct astronomical observation is also difficult because the downstream region often becomes optically thick to its own radiation, preventing it from escaping the system. Thus, only MeV neutrinos and optical radiation can be detected, as shown in Fig. 1. Therefore, the propagation of RS must be studied in laboratories to fully understand the physics underpinning such shocks and the driving mechanism of supernova explosions. Here, we focus on describing the interaction between shocks and envelopes with very different density distributions in progenitor stars by consulting with laboratory RSs.

Many laboratory studies on astrophysics with high-power lasers have already been performed, demonstrating the importance of high-energy-density laboratory astrophysics [6, 7]. Fortunately, RS structures can also be produced in laser facilities, and the physical dynamics properties of plasma materials are similar to those in supernova explosions (Fig. 1).

Various experiments have been conducted and have given valuable highlights on the general structure of RSs. The radiative collapse of the shock front [6] and the radiative precursor [8] have been observed. In parallel, the fundamental parameters of RSs have been measured [9]. RS deceleration has been observed and may contribute to the formation of SN 1987A hotspots [10]. Moreover, experiments on two counter-propagating RSs may be suitable for studies of colliding supernova remnants [11]. However, there has not yet been a scaling connection between laboratory results and astrophysical RS propagation. An indispensable experiment should be done to verify the supernova theory quantitatively.

Here, we present experiments to reproduce the characteristics of CCSNe with different stellar masses and initial explosion energies in a laboratory. In these experiments, shocks are driven in 1.2- and 1.9 atm xenon gas by a laser with energy from 1600 to 2,800 J on the SGIII prototype laser facility. The shock velocity, shocked density, and temperature are obtained in the same experimental shots. With the shock position and recording time, we get the necessary parameters for the first time to scale with CCSNe cases using relevant scaling laws [12]. Furthermore, the rescaled theoretical values are similar to three CCSNe cases with stellar masses of 40 and 50 $M_\odot$ ($M_\odot$ denotes the Sun mass) and initial explosion energies of 1.5 and 2 B (1 B = $10^{44}$ J). Based on the laboratory conditions, the driving mechanism of the supernova explosions could be investigated and unified by multiple cases. These results will contribute to time-domain astrophysical systems, where strong RSs propagate.

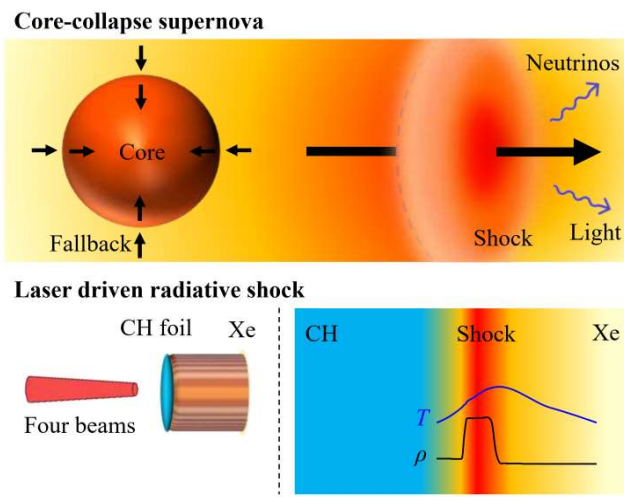

Fig. 1. Similarity in RS propagation between supernova explosions and in the laboratory. At the end of the life of a massive star, the core collapses to a neutron star or a black hole with the following fallback accretion process; meanwhile, a globally relativistic RS runs across the envelope. In the laboratory, high-power laser beams can drive an RS at a velocity higher than 100 km/s in Xe gas. These two shock propagations show that high-speed shocks interact and heat the surrounded gas and that the spatial distributions of density and temperature are physically similar in these two cases. The experiments give us a chance to test supernova models and image the supernova's shocks, which are deeply hidden in the progenitors.

RS was generated on the SGIII prototype laser facility (see Supplementary Material A). The targets were configured as a fistular cuboid with a polyimide tube center. Four laser beams irradiated the CH (styrene at 1.05 g/cm$^3$) disc and accelerated the CH to produce a high-velocity piston, driving a shock wave down the Xe tube. Then, 13 or 15 ns later, the backlight beams illuminated a V foil. The emitted X-rays then passed through the tube, and a monochromatic backlight imaging system (MBIS) obtained the X-ray radiographs at 5.2 keV, derived from the He$_\alpha$ emission of the V source [13]. Fig. 2(a) shows a typical X-ray backlighting radiograph measured by MBIS. This radiograph was observed at 13 ns after the initial laser pulse began. Darker regions were areas of lower signal transmission due to high opacity. The CH layer was positioned to the left of the shocked Xe but was transparent at the 5.2 keV X-ray energy.

We simulated this experiment using the Icefire code developed at the Research Center of Laser Fusion in China. Icefire is a two-dimensional, Lagrangian radiation hydrodynamic code [14]. The simulations accounted for shot-to-shot experimental variations in both the targets and laser energy. Fig. 2(b) displays the simulated radiograph under identical conditions, as depicted in Fig. 2(a). The simulated shock position was consistent with the experimentally observed results at the corresponding time.

The average shock velocity could be obtained by the shock front position $x_{\text{Xe}}$ and $t_{\text{Xe}}$, where the starting point was set as the initial interface of CH and Xe and the starting time was set to the moment the shock entered the Xe gas (see Supplementary Material B). In the experiment, the shocks were driven under different laser energy and Xe pressure conditions, with corresponding photographs captured at 13 or 15 ns. Fig. 2(c) shows the average shock velocities observed under these experimental configurations. Notably, the average shock velocity was above 100 km/s under all the experimental conditions. The average velocity derived from the simulation (red star dots) aligned closely with the experimental results and fell within the margin of error associated with the experimental data. Whether at an Xe pressure of 1.2 atm (black square dots) or 1.9 atm (blue circle dots), the shock velocities increased with the laser energy in the 1500–2800 J range. The influence of Xe pressure on the shock velocity could be assessed at similar input laser energie.

The Xe in the shock tube was compressed into the darker region, and the approximate average postshock density $\rho$ was determined by the compression ratio $R_{\text{com}} = \rho/\rho_0$, with $\rho_0$ as the initial unshocked density. The apparent compression $R_{\text{com}}$ could be obtained by $x_{\text{Xe}}/W_{\text{Xe}}$, where $x_{\text{Xe}}$ is the shock position and $W_{\text{Xe}}$ is the shocked layer width (see Supplementary Material C). From the experimental images, the shocked layer was compressed to above 20 because of the strong radiative cooling effects. Furthermore, the average density of the shocked layer could be derived from $R_{\text{com}}\rho_0$ in the experiments, as shown in Fig. 2(d). The shocked densities were above 0.2 g/cm$^3$ in all cases. The cases with 1.9 atm Xe had higher shocked densities because of the higher $\rho_0$. The two-dimensional (2D) density distributions (Fig. 2(e)) in the tube could be obtained by Icefire code. Thus, we can calculate the average density of the shocked layer within the 60 μm radial center width. The densities obtained through the experiments and the simulations were similar.

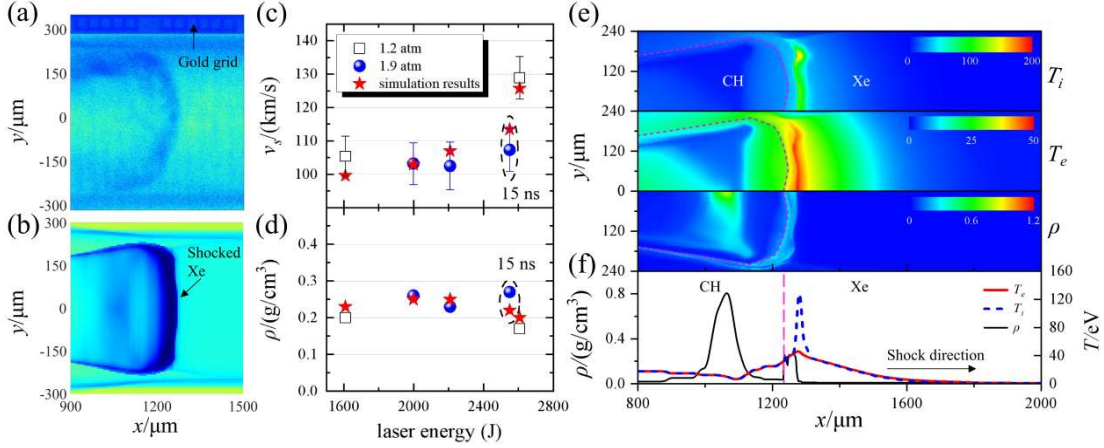

Fig. 2. Experimental results. (a) Typical X-ray backlighting radiograph measured by MBIS at 13 ns. The shocked Xe is labeled and is moving from left to right. The right edge of the dark layer is the shock front, situated at ~1278 μm. The gold grid at the upper part of the image is a spatial fiducia, with squares measuring 42.3 μm per side. (b) Simulated output of a 5.2 keV X-ray backlighting radiograph at 13 ns with the 2D Icefire code. The shock position is about 1,274 μm away from the interface between CH and Xe. (c) Average shock velocity and (d) average shocked density with laser energy and Xe gas pressure. 1.2 atm Xe: black square dots; 1.9 atm Xe: blue circle dots; simulation results: red star dots. The dashed-ellipse marks the data at 15 ns, and the rest are data at 13 ns. (e) Simulated ion temperature (top, color bar: 0–200 eV), electron temperature (middle, color bar: 0–50 eV), and mass density (bottom, color bar: 0–1.2 g/cm$^3$) at 13 ns. The red dash lines represent the interface of CH and Xe. (f) Axial profiles of mass density $\rho$ (black solid line), electron temperature $T_e$ (red dashed line), and ion temperature $T_i$ (blue dash-dotted line) from 2D simulations of RS at 13 ns. The position zero is the initial boundary between CH and Xe. The pink dashed line represents the interface of CH and Xe at 13 ns.

Axial profiles from 2D numerical simulations are shown in Fig. 2(f). The dip in mass density at ~1230 μm marks regions of CH and Xe at 13 ns, whereas the peak in ion temperature of $T_i \approx 125$ eV at ~1280 μm marks the position of the shock front. One can see a narrow, shocked xenon layer due to cooling. This density shows a maximum compression of a factor of ~25 over the initial gas density. At the shock front, the ion temperature was significantly higher than the electron temperature, caused by shock heating. The temperature decreased away from the shock front in the postshock-cooled region. A precursor with about a 500 μm length was generated ahead of the shock. The radiative precursor had a peak temperature of $T_e \approx 46$ eV, which decreased as the propagation distance increased. Furthermore, we could obtain the spatial average temperature using the temperature spatial distribution, which was about 10 eV.

Apparently, mathematical models governing radiation-hydrodynamics-driven phenomena are invariant under homothetic group transformation and can be rescaled according to scaling laws (see Supplementary Material D). The spatial and time dimensionless ratios of astrophysical objects to laboratory plasmas are denoted as $A_x = x/\bar{x}$, $A_t = t/\bar{t}$, where $x$, $t$ express astrophysical objects and $\bar{x}$, $\bar{t}$ express laboratory plasmas. We take $A_x$ and $A_t$ as free scaling factors. For an optically thick RS, the density, velocity, and temperature scaling factors can be expressed as $A_\rho = \frac{\rho}{\bar{\rho}} = \frac{A_x^{11/3}}{A_t^4}$, $A_v = \frac{v}{\bar{v}} = \frac{A_x}{A_t}$, and $A_T = \frac{T}{\bar{T}} = \frac{A_x^2}{A_t^2}$. Then, we can get the rescaled values $\rho$, $v$, and $T$ using the laboratory plasma parameters $\bar{\rho}$, $\bar{v}$, and $\bar{T}$. However, we get the average shock velocities $\bar{v}$ and shocked densities $\bar{\rho}$ from the experiment, and temperature $\bar{T}$ from the simulation. We can thus cast the experiments in the context of CCSNe where a strong RS propagates outward from the center of the star after core collapse.

Here, we adopted the results of spherically symmetric CCSNe explosion simulations [15]. Three CCSNe cases with different stellar masses and initial explosion energies (the metallicities of the progenitor stars were zero) were compared with the experimental cases (see Table S2 in Supplementary Material D). Following the scaling law, three laboratory cases with initial Xe gas densities and laser energies of 1.2 atm and 1,600 J, 1.2 atm and 2,600 J, and 1.9 atm and 2,000 J were rescaled to three CCSNe cases with stellar masses and initial explosion energies of 40 $M_\odot$ and 1.5 B, 40 $M_\odot$ and 2 B, and 50 $M_\odot$ and 2 B, as shown in lines 4, 7, and 10 in Table S2, respectively. Although the temperature was only the same order of magnitude, we had good agreement for the radius, time, and velocity and a small difference between the mass densities in all three cases. In the scaling laws, higher Xe gas densities mean larger stellar masses, and higher laser drive energies mean stronger supernovae explosion energies.

Accompanying the death of a massive star, a drastic supernova explosion might occur, and a neutron star or black hole should be formed in the center. However, after related explorations in the past half century, the explosion mechanism still has great complexity, especially as its key driver is still unclear. Whatever the driving mechanism, a universal picture where a relativistic shock propagates into the envelope is well recognized. Therefore, by setting specific experimental parameters, we can reproduce CCSNe with different stellar masses and explosion energies in the laboratory to study its association with the initial burst energy. Moreover, we can obtain the evolution of the explosion by adjusting the recording time. These can be very useful in deducing the explosion mechanism.

In summary, this work studied the connection between laboratory results and CCSNe concerning scaling laws. We presented RS experiment results performed on an SGIII prototype laser facility with different laser energies and Xe gas densities. From the X-ray backlit transmitted photographs, we obtained the average shock velocities and shocked density. The velocities were higher than 100 km/s, whereas the density compressions were higher than 20 times. Simulations accurately reproduced the experimental results. Furthermore, the temperature of the precursor was obtained through simulation, consistent with the theoretical model. Based on the experimental results, three CCSNe cases with different stellar masses and initial explosion energies were rescaled. The rescaled values had good agreement for radius, time, velocity, and densities in all cases. This study encourages us to continue investigations of CCSNe with different initial masses and explosion energies, based on the connections with laboratory conditions we establish here. We intend for future experiments to get similar characteristics of CCSNe and derive the time evolution of explosions to verify the shock physics in astrophysical phenomena and deepen our understanding of the CCSNe explosion mechanism.

**Acknowledgments**

We acknowledge LFRC staff for operating the laser facility, providing diagnostics and target fabrication. We also thank Chuang Xue from the Institute of Applied Physics and Computational Mathematics for valuable discussions. This work was supported by the National Natural Science Foundation of China (Nos.12335015, 12375238, 12173031, 12303049, and 12105269).

**Data availability**

The data that support the findings of this study are available from the first author upon reasonable request.

**Author contributions**

Lu Zhang proposed and took charge of the radiative shock experiments on the SGIII prototype laser facility. Jianhua Zheng carried out experiments with Longyu Kuang, Shuai Zhang, Longfei Jing, Zhiwei Lin, Liling Li, Hang Li, Jinhua Zheng Zhibing He, and Ping Li. Zhenghua Yang and Pin Yang recorded the X-ray radiographs using MBIS. Yongkun Ding, Tianming Song, Yuxue Zhang, Zhiyu Zhang, Yang Zhao, Dong Yang, Jiamin Yang, and Zongqing Zhao participated in the analysis of experimental results. Longyu Kuang provided the simulation results to account for shot-to-shot experimental variations. Tong Liu and Yunfeng Wei evaluated the properties of the CCSNe explosion. Lu Zhang, Tong Liu, and Yongkun Ding wrote the paper. All authors contributed to the discussions and approved the final version of the manuscript.

**References**

[1] Orlando S, Bonito R, Argiroffi C, et al., *Radiative accretion shocks along nonuniform stellar magnetic fields in classical T Tauri stars.* Astron. Astrophys, 2013: **559**: A127.

[2] Kato S, Fukue J, and Mineshige S, *Black-Hole Accretion Disks: Towards a New Paradigm*. 2008: Kyoto, Japan: Kyoto University Press. 549.

[3] Shapiro SL and Salpeter EE, *Accretion onto neutron stars under adiabatic shock conditions.* Astrophys J, 1975: **198**: 671-682.

[4] Rees MJ and Mészáros P, *Relativistic fireballs: Energy conversion and time-scales.* Mon Not R astr Soc, 1992: **258**: 41-43.

[5] Orlando S, Miceli M, Pumo ML, et al., *Supernova 1987a: A Template to Link Supernovae to Their Remnants.* Astrophys J, 2015: **810**(2): 168.

[6] Reighard AB, Drake RP, Dannenberg KK, et al., *Observation of collapsing radiative shocks in laboratory experiments.* Phys Plasmas 2006: **13**: 082901.

[7] Ping Y, Zhong J, Wang X, et al., *Turbulent magnetic reconnection generated by intense lasers.* Nat Phys, 2023: **19**: 263-270.

[8] Bouquet S, Stehle C, Koenig M, et al., *Observation of laser driven supercritical radiative shock precursors.* Phys Rev Lett, 2004: **92**(22): 225001.

[9] Dizière A, Michaut C, Koenig M, et al., *Highly radiative shock experiments driven by GEKKO XII.* Astrophys Space Sci 2011: **336**(213-218): 213-218.

[10] Michel T, Albertazzi B, Mabey P, et al., *Laboratory Observation of Radiative Shock Deceleration and Application to SN 1987A.* Astrophys J, 2020: **888**(1): 25.

[11] Suzuki-Vidal F, Clayson T, Stehlé C, et al., *Counterpropagating Radiative Shock Experiments on the Orion Laser.* Phys Rev Lett, 2017: **119**: 055001.

[12] Bouquet S, Falize E, Michaut C, et al., *From lasers to the universe: Scaling laws in laboratory astrophysics.* High Energ Dens Phys, 2010: **6**: 368-380.

[13] Wang F, Jiang S, Ding Y, et al., *Recent diagnostic developments at the 100 kJ-level laser facility in China.* Matter Radiat Extremes, 2020: **5**(3): 035201.

[14] Zheng J, Kuang L, Jiang S, et al., *Mitigating wall plasma expansion and enhancing x-ray emission by using multilayer gold films as hohlraum material.* Nucl Fusion, 2021: **61**(8): 086004.

[15] Wei YF, Liu T, and Xue L, *Anisotropic neutrinos and gravitational waves from black hole neutrino-dominated accretion flows in fallback core-collapse supernovae.* Mon Not R astr Soc, 2021: **507**(1): 431-442.

# Supplementary material for "Laboratorial radiative shocks with multiple parameters and first quantifying verifications to core-collapse supernovae"

Lu Zhang[a], Jianhua Zheng[a], Zhenghua Yang[a], Tianming Song[a], Shuai Zhang[a], Tong Liu[b,*], Yunfeng Wei[b], Longyu Kuang[a,*], Longfei Jing[a], Zhiwei Lin[a], Liling Li[a], Hang Li[a], Jinhua Zheng[a], Pin Yang[a], Yuxue Zhang[a], Zhiyu Zhang[a], Yang Zhao[a], Zhibing He[a], Ping Li[a], Dong Yang[a], Jiamin Yang[a], Zongqing Zhao[a], Yongkun Ding[c,*]

[a] Research Center of Laser Fusion, China Academy of Engineering Physics, Mianyang 621900, China

[b] Department of Astronomy, Xiamen University, Xiamen, Fujian 361005, China

[c] Institute of Applied Physics and Computational Mathematics, Beijing 100088, China

**A.** Experimental setup

We use the SGIII prototype laser facility to generate RS. As shown in Fig. S1, the targets are configured as a fistular cuboid with a polyimide tube center. The inside diameter of the tube is 500 μm, and its wall thickness is 25 μm. A gold cirque, with an inside diameter of 650 μm and a thickness of 50 μm, is affixed to the upper surface to retard the progress of the laser-driven shock at radii outside the tube. The drive disks attached to the gold cirques are a 45 μm thick CH (Styrene at 1.05 g/cm$^3$) disc. An air storage chamber linked with the tube is situated beneath the surface, filled with Xenon at pressure 1.9 atm ($\rho$ = 10.4 mg/cm$^3$) or 1.2 atm ($\rho$ = 6.6 mg/cm$^3$).

Four laser beams irradiate CH disk with 0.35 μm wavelength light. The duration of the laser pulse is 1 ns, with about 2.5 kJ of energy on an 800 μm diameter spot. The laser ablates the CH disk and drives a shock wave within it, and then accelerates the CH to produce a high-velocity piston that will drive a shock wave down the Xe tube. 13 ns or 15 ns later, the backlighter beams illuminate a V foil with a diameter of 200 μm. The backlighter beam has a spot diameter of 300 μm and a pulse width of 500 ps.

The emitted X-rays then pass through the tube, and a monochromatic backlight imaging system (MBIS) obtains the X-ray radiographs at 5.2 keV, which is derived from He$_\alpha$ emission of V source [1]. The MBIS takes backlit transmitted photographs of the shock using a spherically bent crystal. The spherically bent crystal is usually set 135 mm away from the sample and provides a magnification of 11.8 in the meridional plane.

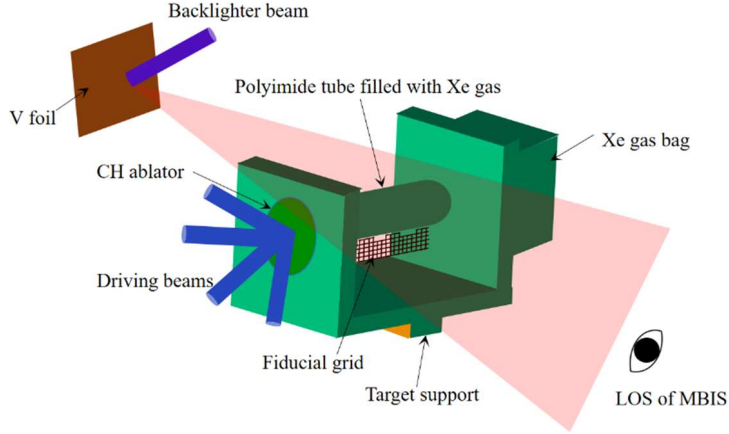

Fig. S1(online). Experimental setup and the target used for the radiative shock created.

**B.** Average shock velocity

Laser ablation of the CH layer initiates a shock, propelling the CH layer to be a high-velocity piston. This CH piston subsequently drives a shock wave down the Xe tube. Fig. S2 shows a typical X-ray backlighting radiograph measured by MBIS. The gold grid fiducial is used to infer a target coordinate system over the experimental image. Darker regions are areas of lower signal transmitted due to high opacity. The CH layer is positioned to the left of the shocked Xe, but is transparent at the 5.2 keV X-ray energy. This radiograph was observed at 13 ns after the initial laser pulse began. The shock is moving from left to right within the tube, with its walls expanding behind it.

The 5.2 keV X-ray transmission at the tube center is shown in Fig. S2 with the red line. The dense layer represents the shocked Xe, where the transmission value is lower than that of the unshocked Xe gas. We set the position of half the max transmission value as the shock front position $x_{Xe}$. $x_{Xe}$ can be determined by the gold grid at the top image, whose position is marked in advance.

The average shock velocity can be obtained by the shock front position $x_{Xe}$ and propagation time $t_{Xe}$, where the starting point is set as the initial interface of CH and Xe, and the starting time is set to the moment that shock enters the Xe gas. The average shock velocity in Xe is

$$\bar{v}_s = \frac{x_{Xe}}{t_{Xe}}$$

(1)

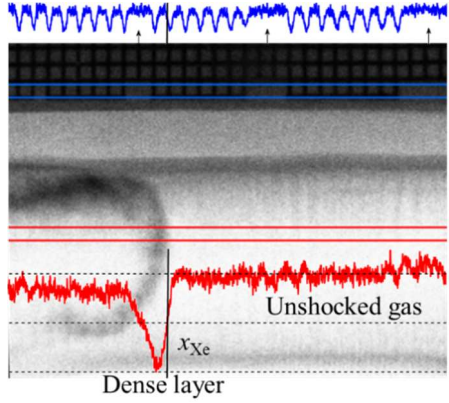

Fig. S2(online). Typical X-ray backlighting radiograph measured by MBIS at 13 ns. The gold grid at the top of the image is a spatial fiducia, with squares 42.3 μm. The blue line shows the transmission of the gold grid, and the red line shows the transmission of Xe gas.

### C. Average density

The darker region in Fig. 2(a) is the dense Xe layer. The right edge of the dark layer delineates the shock front. The Xe in the shock tube is compressed into the darker region, and the approximate average postshock density $\rho$ is determined by the compression ratio of $\rho/\rho_0$, in which $\rho_0$ is the initial unshocked density. To define these measurements quantitatively we utilize lineouts averaged over the tube's lateral dimension. For each image, sufficiently far from the edge effects of the tube wall and wall shock, the radial center 60 μm in width of the shock tube and the axial direction covers the matter ahead of (upstream) the shock and shocked (downstream) region are extracted.

The central lineout of the 13 ns radiograph from Shot 021 is shown in Fig. S3. The counts of image $C_{I0}(x)$-$C_I(x)$ could stand for the absorbed intensity by the tube and the substance in it, where $C_{I0}$ represents an intensity taken outside the tube. There are five regions in $x$, from right to left: 1. Unshocked Xe in the range $x>L_4$; 2. The shock front, where absorbed intensity falls, in the range $L_3<x<L_4$; 3. The dense shocked layer in the range $L_2<x<L_3$; 4. The rising absorbed intensity in the range $L_1<x<L_2$; 5. The transparent region with $x < L_1$, consists mostly of CH[2].

The absorbed intensity primarily depends on the postshock density. Based on the intensity distribution, we can obtain the average shock position $x_{Xe}$, which is $(L_3+L_4)/2$. The shocked Xe gas in the tube is compressed to the shocked layer. Where the shock front intensity of region 2 falls rapidly, has a density jump, and the postshock intensity of region 4 rises slowly. We define the shocked layer width $W_{Xe}$ as the region 3 width plus half of the region 4 and region 2 width, which is $(L_3-L_2)+0.5(L_4-L_3)+0.5(L_2-L_1)$. Then the apparent compression $R_{com}$ is $x_{Xe}/W_{Xe}$. The compressions obtained from the experimental images are shown in Table S1.

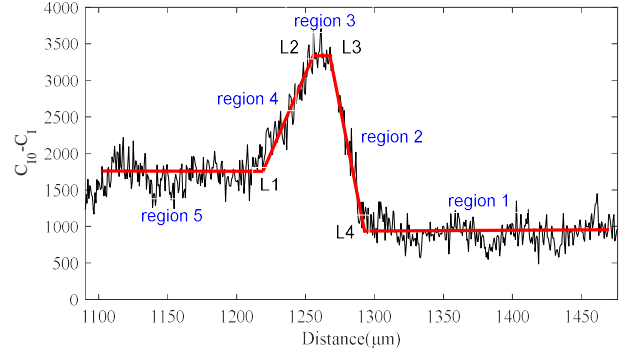

Fig. S3(online). Central lineout of the 13 ns radiograph from Shot 021. showing data averaged (black) over the radial dimension and the five-part piecewise fit to the data (red). Distance from the experiment's drive disc is shown on the *x*-axis (in μm), and the counts of lineout are shown on the *y*-axis.

The shocked layer is compressed to above 20 (radiative collapse) because of the strong radiative cooling effects. Just as the experiment produced thick-downstream, thin-upstream shocks, $R_{com}$ can be expressed as [3]

$$R_{com} = \frac{\rho_f}{\rho_0} = \sqrt{\frac{4Q}{\sqrt{1+8Q}-1}}$$

(2)

Where $\rho_f$ is the shocked density, $Q = 2v_s^5 \sigma/(R^4 \rho_0)$, $R = K_B(Z+1)/(Am_p)$, $K_B = 1.38 \times 10^{-11}$ erg·eV$^{-1}$, is Boltzmann constant, $\sigma = 1.03 \times 10^{24}$ erg·cm$^{-2}$ s$^{-1}$ keV$^{-4}$, is Stefan-Boltzmann constant. The compression is related to the shock velocity and initial density of upstream, which increases with shock velocity $v_s$. However, lower initial upstream density leads to higher compression. As shown in Table S1, the experimental results are consistent with this formula. Specifically, the compression is greater with 1.2 atm Xe compared to 1.9 atm Xe under 45 μm CH disc conditions. This discrepancy is attributed to the lower initial density and higher shock velocity.

Furthermore, the average density of the shocked layer can be derived from $R_{com}\rho_0$ in the experiments, and shown in Table S1. The shocked densities are above 0.2 g/cm$^3$ in all cases. The cases with 1.9 atm Xe have higher shocked densities because of higher $\rho_0$. The 2D density distributions in the tube can be got by Icefire code. So, we can calculate the average density of the shocked layer within the 60 μm radial center width, as presented in Table S1. The densities obtained through experiment and simulation are similar.

Table S1(online). Apparent shock compression ratios, shocked average density by experiments, and the simulated average density using 2D Icefire code for each experimental shot. The critical experimental conditions and the image record times are also indicated.

| Shot number(time) | conditions | Compression by experiments | Shocked Density by experiments(g/cm³) | Shocked Density by simulations(g/cm³) |
|---|---|---|---|---|
| #020(13 ns) | 1.2 atm Xe | 31.0 | 0.20 | 0.23 |
| #027(13 ns) | | 26.5 | 0.17 | 0.2 |
| #021(13 ns) | 1.9 atm Xe | 24.6 | 0.26 | 0.25 |
| #022(13 ns) | | 22.4 | 0.23 | 0.25 |
| #023(15 ns) | | 25.8 | 0.27 | 0.22 |

### D. Scaling laws

The optical depth at frequency $v$ between point $s$ and point $s_0$ is

$$\tau_v = \int_s^{s_0} \chi_v(s')ds'$$

$\chi_v$ is the spectral total opacity in units of $cm^{-1}$, including absorption opacity and scattering opacity. It is optically thick (at some frequency) if $\tau_v \gg 1$, and optically thin if $\tau_v \ll 1$. All materials are optically thick at long enough wavelengths and optically thin at short enough lengths.

In the experiments, shown in Fig. 2(f), the electron temperature of the downstream is about 30 eV, and the density is about 0.2 g/cm³. what of the upstream is about 40 eV and 0.01 g/cm³. The maximal spectral intensity is located at about 100 eV by Planck's formula. By the opacity data of Xe, $\chi_{100eV}$ is 0.31 μm$^{-1}$ at the downstream region, and 0.017 μm$^{-1}$ at the upstream region. Set the distance as the shocked layer width, about 40 μm, then we can get the optical depth of downstream is 12.5, and 0.70 of upstream. So, it is optically thick downstream and optically thin upstream.

The system of equations governing the behavior of radiation hydrodynamics in optically thick plasmas consists of the continuity equation:

$$\frac{\partial}{\partial t}\rho(x,t) + \frac{1}{x^N}\frac{\partial}{\partial x}[x^N \rho(x,t)v(x,t)] = 0 \tag{3}$$

the equation of motion

$$\left(\frac{\partial}{\partial t} + v\frac{\partial}{\partial x}\right)v(x,t) = -\frac{1}{\rho}\frac{\partial}{\partial x}p(x,t) \tag{4}$$

And the energy equation

$$\left(\frac{\partial}{\partial t} + v\frac{\partial}{\partial x}\right)p(x,t) - \gamma\frac{p}{\rho}\left(\frac{\partial}{\partial t} + v\frac{\partial}{\partial x}\right)\rho(x,t) = -(\gamma - 1)\frac{1}{x^N}\frac{\partial}{\partial x}[x^N F(\rho,T)] \tag{5}$$

Where $F(\rho,T)$ is the radiation flux given by

$$F(\rho,T) = -\kappa(\rho,T)\frac{\partial}{\partial x}T(x,t) \tag{6}$$

$N$ is the geometrical exponent ($N = 0$, 1, and 2 in plane, cylindrical and spherical geometry, respectively), $\gamma$ is the polytropic index, and $x$, $t$, $\rho(x,t)$, $v(x,t)$, $p(x,t)$,

$T(x, t)$ stand respectively for space, time, mass density, velocity, pressure and temperature.

The equation of state (EOS)

$$p(\rho, T) = C_{EOS} \rho^\mu T^\nu \quad (7)$$

And the opacity

$$\kappa(\rho, T) = \kappa_0 \rho^m T^n \quad (8)$$

For the study of the invariance of the model governed by (3)- (8), we use the transformation defined by T[4],

T: $(x, t, v, p, \rho, T, F, \kappa, C_{EOS}, \kappa_0) \to (\bar{x}, \bar{t}, \bar{v}, \bar{p}, \bar{\rho}, \bar{T}, \bar{F}, \bar{\kappa}, \overline{C_{EOS}}, \overline{\kappa_0})$

The quantities with an upper bar correspond to the laboratory experiment, and the quantities without an upper bar correspond to the astrophysical object. The values of the scaling factors $A_i = \frac{i}{\bar{i}}$ $(i = x, t, v, p, \rho, T, F, \kappa)$ will make possible the connection between astrophysical objects and laboratory plasmas.

With $A_x$ and $A_t$ free scaling factors, scaling laws obtained for the strict invariance of the equations of the model are:

$$A_\rho = \frac{(A_x)^{\frac{2(2\nu-n-1)}{\nu(m-1)-(n+1)(\mu-1)}}}{(A_t)^{\frac{3\nu-2n-2}{\nu(m-1)-(n+1)(\mu-1)}}}$$

$$A_v = \frac{A_x}{A_t}$$

$$A_p = \frac{(A_x)^{\frac{2\nu(m+1)-2\mu(n+1)}{\nu(m-1)-(n+1)(\mu-1)}}}{(A_t)^{\frac{\nu(2m+1)-2\mu(n+1)}{\nu(m-1)-(n+1)(\mu-1)}}}$$

$$A_T = \frac{(A_x)^{\frac{2(m+1-2\mu)}{\nu(m-1)-(n+1)(\mu-1)}}}{(A_t)^{\frac{2m+1-3\mu}{\nu(m-1)-(n+1)(\mu-1)}}}$$

$$A_F = \frac{(A_x)^{\frac{\nu(3m+1)-(n+1)(3\mu-1)}{\nu(m-1)-(n+1)(\mu-1)}}}{(A_t)^{\frac{3\nu m-(n+1)(3\mu-1)}{\nu(m-1)-(n+1)(\mu-1)}}}$$

$$A_\kappa = (A_x)^{\frac{2[m(2\nu-1)+n(1-2\mu)]}{\nu(m-1)-(n+1)(\mu-1)}} \times (A_t)^{\frac{m(2-3\nu)+n(3\mu-1)}{\nu(m-1)-(n+1)(\mu-1)}}$$

As the temperature is rather high ($T \approx 10$ eV), so we may consider that the main contribution to the opacity of the plasma is the free-free absorption given by Kramers law[5] with $m = -2$ and $n = 13/2$. To perform the connection between the laboratory experiments and SNII, we have assumed that an ideal gas EOS ($\mu = \nu = 1$) is a good approximation since the density in the experiment is low and the temperature in the supernova progenitor is very high.

Then we get the scaling factors as $A_\rho = \frac{\rho}{\bar{\rho}} = \frac{A_x^{11/3}}{A_t^4}$, $A_v = \frac{v}{\bar{v}} = \frac{A_x}{A_t}$, $A_p = \frac{p}{\bar{p}} = \frac{A_x^{17/3}}{A_t^6}$, $A_T = \frac{T}{\bar{T}} = \frac{A_x^2}{A_t^2}$, $A_F = \frac{F}{\bar{F}} = \frac{A_x^{20/3}}{A_t^7}$, $A_\kappa = \frac{\kappa}{\bar{\kappa}} = \frac{A_x^{17/3}}{A_t^5}$ for the connection between the laboratory experiments and SNII.

We think that the initial explosion energies drive the supernova explosion with a certain velocity, and the stellar masses decide the initial density spatial distributions. The RS values of the characteristic length, time, velocity, mass density, and temperature are shown in Table S2. For CCSNe, V is the average shock velocity after the explosion of 100 s, and T is its temperature derived from the local density and pressure. From the laboratory data $\bar{x}$, $\bar{t}$, $\bar{\rho}$, $\bar{v}$, and $\bar{T}$, the theoretical rescaled $x$, $t$, $\rho$, $v$, and $T$ are computed, shown in lines 4, 7, and 10 in Table S2.

Table S2(online). Values of various physical quantities at the shock front in a core-collapse supernova, laser-driven RS, and rescaled values deduced from laboratory data. The stellar masses and initial explosion energies are 40 M☉ & 1.5 B, 40 M☉ & 2 B, and 50 M☉ & 2 B, respectively. Corresponding, the Experimental values with initial Xe densities and laser energies are 1.2 atm & 1600 J, 1.2 atm & 2600 J, and 1.9 atm & 2000 J, respectively.

| | Physical quantities | R(cm) | Time(s) | V (km/s) | Density(g/cm³) | T/K |
|---|---|---|---|---|---|---|
| 40 M☉ & 1.5 B (astrophysics) vs 1.2 atm & 1600 J (laboratory) | CCSNe | $3\times10^{10}$ | 100 | 3000 | 20 | $1.1\times10^8$ |
| | Experimental values | 0.13 | $1.25\times10^{-8}$ | 105 | 0.2 | $4\times10^5$ |
| | Rescaled values 1 | $3\times10^{10}$ | 100 | 3000 | 23 | $3.3\times10^8$ |
| 40 M☉ & 2 B (astrophysics) vs 1.2 atm & 2600 J (laboratory) | CCSNe | $3.5\times10^{10}$ | 100 | 3500 | 21 | $1.1\times10^8$ |
| | Experimental values | 0.16 | $1.25\times10^{-8}$ | 130 | 0.17 | $5\times10^5$ |
| | Rescaled values 2 | $3.5\times10^{10}$ | 100 | 3500 | 16 | $3.7\times10^8$ |
| 50 M☉ & 2 B (astrophysics) vs 1.9 atm & 2000 J (laboratory) | CCSNe | $3\times10^{10}$ | 100 | 3000 | 34 | $1.4\times10^8$ |
| | Experimental values | 0.128 | $1.25\times10^{-8}$ | 100 | 0.26 | $4.5\times10^5$ |
| | Rescaled values 3 | $3\times10^{10}$ | 100 | 2900 | 31 | $3.8\times10^8$ |

# References

[1] Wang F, Jiang S, Ding Y, et al., *Recent diagnostic developments at the 100 kJ-level laser facility in China.* Matter Radiat Extremes, 2020: **5**(3): 035201.

[2] Doss FW, Drake RP, and Kuranz CC, *Statistical inference in the presence of an inclination effect in laboratory radiative shock experiments.* Astrophys Space Sci, 2011: **336**: 219-224.

[3] Drake RP, *High-Energy-Density Physics*. 2006, Netherlands: Springer.

[4] Bouquet S, Falize E, Michaut C, et al., *From lasers to the universe: Scaling laws in laboratory astrophysics.* High Energ Dens Phys, 2010: **6**: 368-380.

[5] Rybicki GB and Lightman AP, *Radiative Processes in Astrophysics*. 2004, Weinheim: WILEY-VCH Verlag GmbH & Co. KGaA.